%% file: main.tex
\documentclass[twocolumn]{aastex701}

\input{preamble}

\graphicspath{ {./} }

\begin{document}

\title{The Squeezed Bispectrum from CHIME HI Emission and Planck CMB Lensing: Current Sensitivity and Forecasts}

\input{authorlist.tex}

\input{abstract}

\input{intro}
\input{theory}
\input{data}
\input{method}
\input{results}
\input{forecasts}
\input{conclusion}
\input{ack}

\bibliographystyle{aasjournalv7}
\bibliography{main}

\end{document}

%% file: preamble.tex
\usepackage{xcolor}
\usepackage{amsmath}
\usepackage{amssymb}
\usepackage{xspace}
\usepackage{tabularx}
\usepackage{graphicx}
\usepackage{bm}

\newcommand{\centfig}[2][1]{
    \centering
    \makebox[\linewidth][c]{\includegraphics[width=#1\linewidth]{#2}}
}

\newcommand{\degree}{$^{\circ}$\xspace}
\newcommand{\tcm}{21$\,$cm\xspace}  
\newcommand{\HI}{\ensuremath{{\rm HI}}\xspace}
\newcommand{\code}[1]{\texttt{#1}}

\newcommand{\lv}{\ensuremath{\left\lvert}}   
\newcommand{\rv}{\ensuremath{\right\rvert}}

\newcommand{\lp}{\ensuremath{\left(}}        
\newcommand{\rp}{\ensuremath{\right)}}

\renewcommand{\vec}[1]{\bm{#1}}

%% file: authorlist.tex
\newcommand{\UBC}{Department of Physics and Astronomy, University of British Columbia, Vancouver, BC, Canada}
\newcommand{\MITP} {Department of Physics, Massachusetts Institute of Technology, Cambridge, MA, USA}
\newcommand{\MITK} {MIT Kavli Institute for Astrophysics and Space Research, Massachusetts Institute of Technology, Cambridge, MA, USA}
\newcommand{\TRU}{Department of Physical Sciences, Thompson Rivers University, Kamloops, BC, Canada}
\newcommand{\PI}{Perimeter Institute for Theoretical Physics, Waterloo, ON, Canada}
\newcommand{\DRAO}{Dominion Radio Astrophysical Observatory, Herzberg Astronomy \& Astrophysics Research Centre, National Research Council Canada, Penticton, BC, Canada}
\newcommand{\UBCO}{Department of Computer Science, Math, Physics, and Statistics, University of British Columbia-Okanagan, Kelowna, BC, Canada}
\newcommand{\McGill}{Department of Physics, McGill University, Montréal, QC, Canada}
\newcommand{\MU}{\McGill}
\newcommand{\TSI}{Trottier Space Institute, McGill University, 3550 rue University, Montréal, QC H3A 2A7, Canada}
\newcommand{\UofTastro}{David A.\ Dunlap Department of Astronomy \& Astrophysics, University of Toronto, Toronto, ON, Canada}
\newcommand{\UTA}{\UofTastro}
\newcommand{\UofTphys}{Department of Physics, University of Toronto, Toronto, ON, Canada}
\newcommand{\WVU} {Department of Computer Science and Electrical Engineering, West Virginia University, Morgantown WV, USA}
\newcommand{\WVUA} {Department of Physics and Astronomy, West Virginia University, Morgantown, WV, USA}
\newcommand{\WVUGWAC} {Center for Gravitational Waves and Cosmology, West Virginia University, Morgantown, WV, USA}
\newcommand{\Yale}{Department of Physics, Yale University, New Haven, CT, USA}
\newcommand{\YUP}{\Yale}
\newcommand{\YaleA}{Department of Astronomy, Yale University, New Haven, CT, USA}
\newcommand{\Dunlap}{Dunlap Institute for Astronomy and Astrophysics, University of Toronto, Toronto, ON, Canada}
\newcommand{\DIAA}{\Dunlap}
\newcommand{\RRI}{Raman Research Institute, Sadashivanagar,   Bengaluru, India}
\newcommand{\ASIAA}{Institute of Astronomy and Astrophysics, Academia Sinica, Taipei, Taiwan}
\newcommand{\CITA}{Canadian Institute for Theoretical Astrophysics, Toronto, ON, Canada}
\newcommand{\CIFAR}{Canadian Institute for Advanced Research,  Toronto, ON, Canada}
\newcommand{\WVUphysastro}{Department of Physics and Astronomy, West Virginia University, Morgantown, WV, USA}
\newcommand{\KIPAC}{Kavli Institute for Particle Astrophysics and Cosmology, Stanford, CA 94305, USA}
\newcommand{\SLAC}{SLAC National Accelerator Laboratory; Menlo Park, CA 94025; USA}
\newcommand{\ASU}{Department of Physics, Arizona State University, Tempe, AZ 85287, USA}
\newcommand{\IAS}{School of Natural Sciences, Institute for Advanced Study, 1 Einstein Drive, Princeton, NJ 08540, USA}

\shortauthors{CHIME Collaboration}

\collaboration{all}{The CHIME Collaboration}

\author[0000-0002-7758-9859]{Arnab Chakraborty}
\affiliation{\MU}
\affiliation{\TSI}
\email{arnab.chakraborty2@mail.mcgill.ca}

\author[0000-0001-7166-6422]{Matt Dobbs}
\affiliation{\MU}
\affiliation{\TSI}
\email{nan}

\author[0000-0002-0190-2271]{Simon Foreman}
\affiliation{\ASU}
\email{simon.foreman@asu.edu}

\author[0000-0003-3986-954X]{Liam Gray}
\affiliation{\UBC}
\email{liam.gray@ubc.ca}

\author[0000-0002-1760-0868]{Mark Halpern}
\affiliation{\UBC}
\email{halpern@physics.ubc.ca}

\author[0000-0002-4241-8320]{Gary Hinshaw}
\affiliation{\UBC}
\email{hinshaw@phas.ubc.ca}

\author[0000-0003-4179-4073]{Albin Joseph}
\affiliation{\ASU}
\email{ajosep52@asu.edu}

\author[0000-0001-8064-6116]{Joshua MacEachern}
\affiliation{\DRAO}
\email{joshua.maceachern@nrc-cnrc.gc.ca}

\author[0000-0002-4279-6946]{Kiyoshi W. Masui}
\affiliation{\MITK}
\affiliation{\MITP}
\email{kmasui@mit.edu}

\author[0000-0002-0772-9326]{Juan Mena-Parra}
\affiliation{\DIAA}
\affiliation{\UTA}
\email{juan.menaparra@utoronto.ca}

\author[0000-0002-7333-5552]{Laura Newburgh}
\affiliation{\YUP}
\email{laura.newburgh@yale.edu}

\author[0000-0002-9516-3245]{Tristan Pinsonneault-Marotte}
\affiliation{\KIPAC}
\affiliation{\SLAC}
\email{tristpm@stanford.edu}

\author[0000-0001-6967-7253]{Alex Reda}
\affiliation{\YUP}
\email{alex.reda@yale.edu}

\author[0000-0001-6731-0351]{Shabbir Shaikh}
\affiliation{\ASU}
\email{sshaik14@asu.edu}

\author[0000-0003-2631-6217]{Seth Siegel}
\affiliation{\PI}
\affiliation{\MU}
\affiliation{\TSI}
\email{ssiegel@perimeterinstitute.ca}

\author[0000-0002-1491-3738]{Haochen Wang}
\affiliation{\MITP}
\affiliation{\MITK}
\email{hcwang96@mit.edu}

\author[0000-0001-7314-9496]{Dallas Wulf}
\affiliation{\MU}
\affiliation{\TSI}
\email{dallas.wulf@mcgill.ca}

\collaboration{all}{And}

\author[0000-0002-9957-0448]{Zeeshan Ahmed}
\affiliation{\KIPAC}
\affiliation{\SLAC}
\email{zeesh@slac.stanford.edu}

\author[0000-0002-5808-4708]{Nickolas Kokron}
\affiliation{\IAS}
\email{kokron@ias.edu}

\author[0000-0002-4619-8927]{Emmanuel Schaan}
\affiliation{\KIPAC}
\affiliation{\SLAC}
\email{eschaan@stanford.edu}

\correspondingauthor{Tristan Pinsonneault-Marotte}
\email{tristpm@stanford.edu}

\correspondingauthor{Emmanuel Schaan}
\email{eschaan@stanford.edu}

\correspondingauthor{Zeeshan Ahmed}
\email{zeesh@slac.stanford.edu}

%% file: abstract.tex
\begin{abstract}
    Line intensity mapping using atomic hydrogen (\HI) has the potential to
    efficiently map large volumes of the universe if the signal can be
    successfully separated from overwhelmingly bright radio foreground emission.
    This motivates cross-correlations, to ascertain the cosmological nature of measured HI fluctuations, and to study  their connections with galaxies and the underlying matter density field. 
    However, these same foregrounds render the cross-correlation with projected fields such as the lensing of the cosmic microwave background (CMB) difficult.
    Indeed, the correlated Fourier modes vary slowly along the line of sight,
    and are thus most contaminated by the smooth-spectrum radio continuum foregrounds.
    In this paper, we implement a method that avoids this issue by
    attempting to measure the non-linear gravitational coupling of the small-scale \tcm power from the Canadian Hydrogen Intensity Mapping
    Experiment (CHIME) with large-scale \textit{Planck} CMB lensing. 
    This measurement is a position-dependent power spectrum, i.e. a squeezed integrated bispectrum.
    Using $94$ nights
    of CHIME data between $1.0 < z < 1.3$ and aggressive foreground filtering,
    we find that the expected signal is five times smaller than the current noise.
    We forecast that incorporating the additional nights of CHIME data already collected would enable a signal-to-noise ratio of 3, without any further improvements in filtering for foreground cleaning.
    
\end{abstract}

%% file: intro.tex
\section{Introduction}

Using the \tcm line of neutral Hydrogen (\HI) to perform intensity mapping is a
technique that has the potential to probe large cosmological volumes out to
redshifts where other tracers are sparse or non-existent
\citep{chang_2008,ansari_2019}. In the last decade, dedicated experiments have
been built to perform this measurement, including CHIME at redshifts targeting
the onset of dark energy \citep{chime_overview_2022} and HERA for reionisation
\citep{hera_2016}. A number of general-purpose instruments have made targeted
observations aimed at measuring this signal, including the GBT, Parkes, LOFAR,
LWA, MWA, GMRT, and MeerKAT. Detections have been reported in cross-correlation with
spectroscopic galaxy surveys
\citep{pen_2009,chang_deep2_2010,gbt_xcorr_2013,anderson_parkes_2018,tramonte2020,li2021-parkesxwigglez,wolz2021,chime_collaboration_detection_2022,chen2025-meerkat}
and the Lyman-$\alpha$ forest \citep{chime_lya_2024}. Measurements of the power
spectrum at $z=0.32$ and $z=0.44$ have been claimed by \citet{paul_2023}, and more recently
\citet{chime_ps_2025} have reported a detection at $z\sim 1$. At high redshift, upper
limits are regularly being improved \citep{hera_ps_2022,gmrt_2021}. Hydrogen
intensity mapping is a core science driver for upcoming radio instruments like
CHORD \citep{chord_2019} and next-generation facilities like the SKA
\citep{ska_cosmo_2020}.

Despite the significant progress in recent years, intensity mapping using the
\tcm line remains a challenging measurement. Extremely bright foreground
emission from our galaxy and extragalactic point sources, in addition to an
ever-worsening RFI environment, obscure the faint cosmological signal. Even as
analysis methods continue to improve and internal detections gain in
significance, cross-correlations will be sure to play a role as powerful
cross-checks for the robustness of these results and to provide extra
constraining power by breaking degeneracies or accessing additional information.

As cited above, galaxy catalogs have provided an excellent target for high-significance cross-correlation measurements. At redshifts $z \gtrsim 2$ however, such surveys become more difficult or impossible to perform from the ground due to the reduced atmospheric transmission at longer wavelengths. Hydrogen intensity mapping on the other hand is poised to observe at these higher redshifts, so identifying alternative surveys that have overlap for cross-correlation will be useful. One compelling possibility is to use lensing of the CMB by structures along the line of sight. The lensing signal includes contributions from high redshift, which should be correlated with corresponding \tcm measurements. As a probe of large-scale structure, lensing is a very clean tracer of the matter density, which would allow us to probe the connection between \tcm fluctuations and the underlying matter density field.

\begin{figure*}[htb]
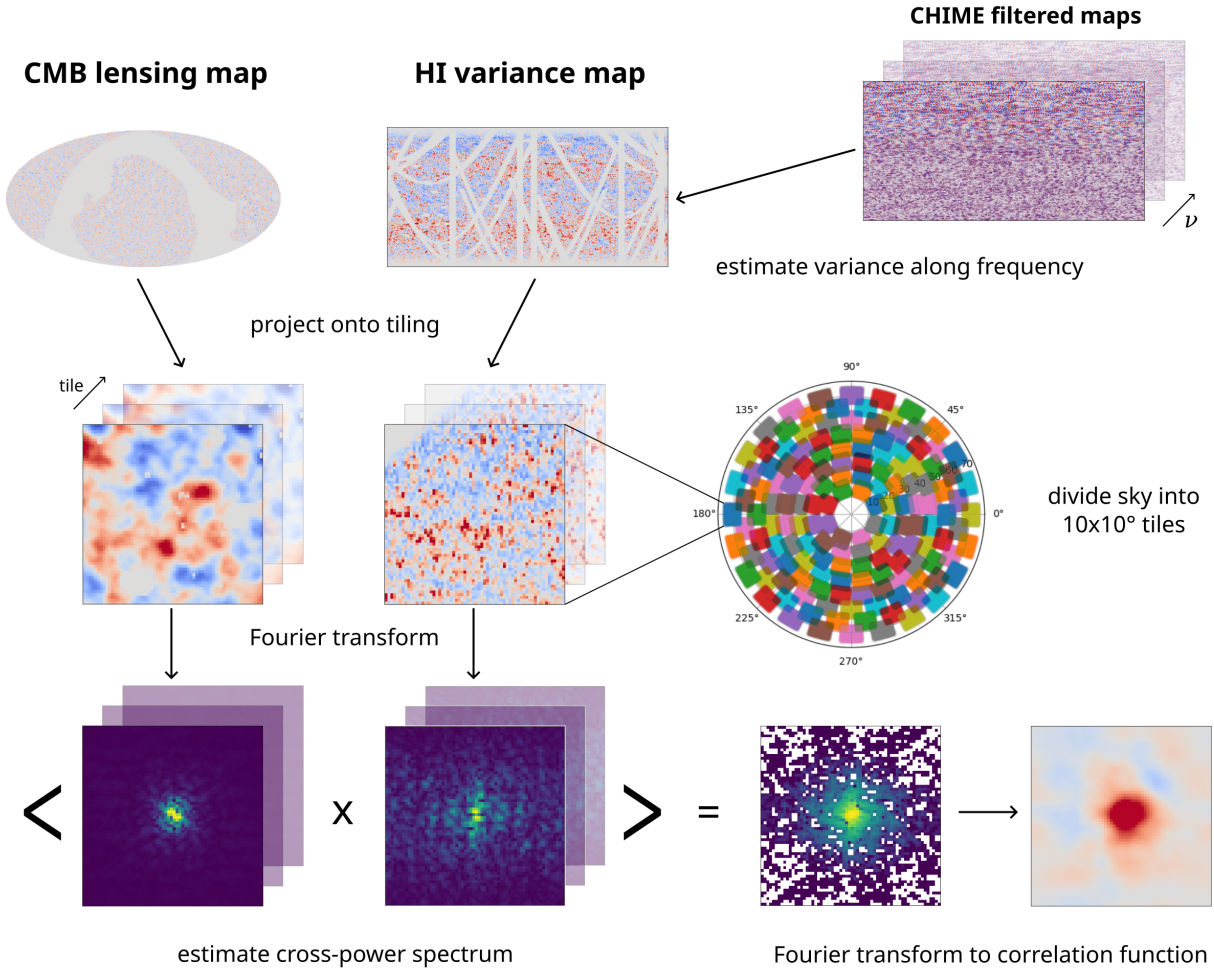

    \centfig[0.95]{drawing}
    \caption{We estimate the cross-correlation of the lensing convergence and the variance of the line-of-sight neutral hydrogen fluctuations
    $\langle \HI^2\rangle$ fields using a tiled flat-sky approach. 
    Noise-free non-Gaussian simulations of the expected signal from \texttt{SkyLine} \citep{skyline_2023} were used to produce the maps shown here.}
    \label{fig:diagram}
\end{figure*}

At first glance this cross-correlation seems like a non-starter: the smooth
modes along the line of sight that are measured by lensing are precisely those
that are maximally contaminated by smooth-spectrum foregrounds in the \tcm
observations, and must be filtered out to access the cosmological signal. 
This
makes a direct two-point cross-correlation between these measurements
appear infeasible. 
However, nonlinear effects that couple power at small and large
scales may enable its detection in higher-order statistics. 
Several such nonlinear effects exist.
On the one hand, nonlinear coupling of line-of-sight Fourier modes occurs due to the redshift-evolution of the line luminosity \citep{delon_2025}.
At the CHIME level of sensitivity and foreground removal however, \citet{delon_2025} have shown that this effect is not sufficient to make a direct cross-correlation detectable.
Another form of nonlinear coupling is from gravitational lensing, e.g. \citet{simon_lensing_2018, manu_cib_2018}.
Finally, the nonlinear evolution due to gravity provides another mechanism for this coupling.
Recovering lost modes in the \tcm maps has been proposed using tidal reconstruction \citep{zhu_recovering_2016,li-ksz-tidal_2019,modi_2019,karacayli-tidal_2019,darwish_density_2021,zhu-tidal-halos_2022,zang-tidal-zspace_2024}, or
by targeting a higher-order statistic such as the bispectrum
\citep{chiang_position-dependent_2014}. 
The latter is the method we will adopt
in this work, by cross-correlating a map of the line-of-sight variance of the
\tcm field observed by CHIME with a map of CMB lensing convergence from Planck.
As we will demonstrate in the following section, the result is an estimate of
the integrated bispectrum. The procedure is summarised in
Figure~\ref{fig:diagram}, and details are provided in Section~\ref{sec:method}.

A detection of the bispectrum using a very similar method applied to CMB lensing
and the Lyman-$\alpha$ forest was first reported by \citet{doux_first_2016}.
Another measurement has recently been published in cross-correlation with DESI
quasars \citep{lya_bispec_2024}, and the sensitivity to this effect is forecast
to improve greatly with upcoming surveys \citep{laposta_lya_2025}.
For \tcm,
\citet{moodley_cross-bispectrum_2023} have explored this approach using a Fisher
analysis.
They forecast improved constraints on cosmological
parameters when the cross-correlation information is added to an auto-spectrum
detection.

In this work we focus on assessing the feasibility of a detection, and will use simulations propagated through instrumental processing to forecast the signal, rather than an analytical calculation. For illustration however, we sketch out a derivation of the theoretical expectation for this
cross-correlation in Section~\ref{sec:theory}. 
We describe the data used for the analysis in
Section~\ref{sec:data}, explain the cross-correlation method in
Section~\ref{sec:method}, and discuss the result in Section~\ref{sec:results}.
Finally, we explore the possibility of future detections using simulation-based
forecasts in Section~\ref{sec:forecasts}.

%% file: theory.tex
\section{Theory}
\label{sec:theory}

\subsection{A Measurement of the Squeezed Bispectrum}

To gain some intuition about what we aim to measure, we briefly sketch out the
relation between the lensing convergence / \HI variance cross-correlation and
the underlying density field. Consider a map of \tcm emission measured as a
function of frequency $\nu$ and position $\vec{\theta}$ on a small, flat patch
of sky: $\delta T(\vec{\theta}, \nu)$. The \tcm fluctuations trace the matter
overdensity field via a mapping of angular position and frequency to
3-dimensional comoving space, modulated by a bias factor that we will assume is
simply linear, i.e.
 \begin{equation}
    \delta T(\vec{\theta}, \nu) = b_\HI \delta_m(\chi(\nu)\vec{\theta}, \chi(\nu))\ .
 \end{equation}
 We measure the variance along the line of sight by summing the squared maps
over the frequency axis:
\begin{equation}
    \delta T^2(\vec{\theta})
    \equiv \int d\nu \delta T^2(\vec{\theta}, \nu)
    = \int d\chi W_\HI(\chi) \delta_m^2(\chi \vec{\theta}, \chi)\ ,
\end{equation}
where the $\int d\chi W_\HI(\chi)$ encodes the conversion of the sum over
frequency into an integral over line of sight distance $\chi$, and absorbs the
linear \HI bias factor.
In
reality, the processing of the maps will lead to a more complicated expression
--- notably, the filtering of foregrounds leads to a mixing between frequencies
--- but for simplicity in this derivation we ignore such details. In the
simulation described later all of these effects are present. In the flat sky
approximation, the angular Fourier transform of this map is
\begin{equation}
    \delta T^2(-\vec{\ell}) = \int d\chi W_\HI(\chi) \int dk_\perp^2
    \delta_m(\vec{k}_\perp, \chi) \delta_m^*(\vec{k}_\perp + \vec{\ell} / \chi, \chi)\ ,
\end{equation}
a statement that the squared map is a convolution of the density field
with itself in Fourier space.

Similarly, given the lensing kernel $W_\kappa$, the map of CMB lensing convergence can be expressed as:
\begin{equation}
    \kappa(\vec{\ell}) = \int d\chi W_\kappa(\chi) \delta_m(\vec{\ell} / \chi, \chi)\ .
\end{equation}
The quantity we aim to compute in this work is the angular cross-correlation of the
two,
\begin{widetext}
    \label{eq:xcorr-model}
\begin{equation}
    \langle \delta T^2_{- \vec{\ell}} \kappa_{\vec{\ell}}\rangle
    = \int d\chi d\chi' W_\HI(\chi) W_\kappa(\chi') \int dk_\perp^2
    \langle \delta_m(\vec{k}_\perp, \chi) \delta_m^*(\vec{k}_\perp + \vec{\ell} / \chi, \chi)
    \delta_m(\vec{\ell} / \chi', \chi')\rangle\ .
\end{equation}
\end{widetext}
In order to interpret this expression, we need to understand how the respective
line of sight projection integrals interact. In the case of the \HI part, the
kernel function $W_\HI$ has support over only a relatively narrow range of
$\chi$, corresponding to the bandwidth of the redshifted \tcm observations, and
is slowly varying compared to the density fluctuations. Evaluated within this
shell, the integral of power over $\chi$ can be cast in terms of Fourier modes
along the line of sight (Plancherel theorem),
\begin{equation}
\begin{aligned}
    \int d\chi W_\HI(\chi) \delta_m(\vec{k}_\perp, \chi)
    \delta_m^*(\vec{k}_\perp + \vec{\ell} / \chi, \chi)&\\
    \approx W_\HI(\chi_0)\int dk_\parallel \delta_m^L(\vec{k}_\perp, k_\parallel)
    \delta_m^{L*}(\vec{k}_\perp + \vec{\ell} / \chi_0, k_\parallel)&\ ,
\end{aligned}
\end{equation}
where the $L$ superscript reminds us that these are modes that belong to the
shell locally centered on $\chi_0$. For illustration, we have ignored the effect
of $W_\HI$ on the line of sight integral, which would lead to mixing of
$k_\parallel$ modes (e.g.\ \citealt{delon_2025}), and we have neglected the
evolution of $\chi$ across the shell, approximating the projection effect by the
central value: $\vec{\ell} / \chi_0$.

The second line of sight integral corresponding to the convergence term will
only pick up correlations between the $\delta_m$ within the same shell centered
on $\chi_0$ where they are coherent. Over this range, the kernel $W_\kappa$ is
roughly constant, so this integral results in an average of $\delta_m$ over the
local shell:
\begin{equation}
\begin{aligned}
    \int d\chi' W_\kappa(\chi') \delta_m(\vec{\ell} / \chi', \chi')&\\
    \approx\ W_\kappa(\chi_0)\delta_m^L(\vec{\ell} / \chi_0, k_\parallel = 0)&\ ,
\end{aligned}
\end{equation}
i.e. the $k_\parallel = 0$ mode.

Putting this all together, we have an expression for the cross-correlation of
the lensing convergence and the \tcm variance maps:
\begin{widetext}
\begin{equation}
    \langle \delta T^2_{- \vec{\ell}} \kappa_{\vec{\ell}}\rangle
    \propto \int dk_\perp^2 dk_\parallel\ 
    \langle\delta_m^L(\vec{k}_\perp, k_\parallel)
    \delta_m^{L*}(\vec{k}_\perp + \vec{\ell} / \chi_0, k_\parallel)
    \delta_m^L(\vec{\ell} / \chi_0, k_\parallel = 0)\rangle\ ,
    \label{eq:xcorr-th}
\end{equation}
\end{widetext}
where we've dropped the overall normalisation factors outside of the integral.
In this form, it is clear that this product is related to an integral over
components of the bispectrum, evaluated within the shell where the \HI
observations are present. The $k_\parallel$ that contribute to the sum will be
on the scale of $\sim 2\pi / \delta\chi$, the line of sight resolution of the
\tcm maps, whereas the transverse modes are on the scale $\ell / \chi_0$. For
the CHIME maps at the telescope's native frequency channel width, these numbers
are respectively on the order of $2\pi / \delta\chi \sim 2 \ \mathrm{Mpc}^{-1}$
and $(\ell = 100) / \chi_0\sim 0.02\ \mathrm{Mpc}^{-1}$. The disparity
$k_\parallel \gg k_\perp$ indicates that this quantity is sensitive to squeezed
modes of the bispectrum, as expected since it is closely related to the
position-dependent power spectrum described by
\citet{chiang_position-dependent_2014}. In that paper, the authors have shown
that the integrated bispectrum in this configuration to leading order is given
by the response of the small-scale power spectrum to large-scale density
fluctuations:
\begin{equation}
    \langle\delta T^2_{-\vec{\ell}} \kappa_{\vec{\ell}}\rangle \propto \sigma_L^2 \
    \frac{d P(\ell/\chi_0)}{d\delta^L_m} \ ,
    \label{eq:pps}
\end{equation}
where $\sigma^2_L = \int d^3k P(k)$ is the variance of the density field within
the local volume.

The sketch above provides some insight into the measurement, but turning it into
a model with predictive power and physical parameters will require expanding the
bispectrum or the form in Equation~\ref{eq:pps}, as well as a careful accounting
of the redshift space effects and bias for the \HI field. 
At quasi-linear scales, perturbation theory could be used to predict the relevant
bispectrum \citep{darwish_density_2021,karagiannis_21cmbispectrum_2022}, while at nonlinear scales, one would need to resort to
\HI halo models \citep{padmanabhan_hihalomodel_2017,wolz_hihalomodel_2019,schaan_IMhalomodel_2021,chen_hihalomodel_2021} or simulation-based predictions 
\citep{modi_hiddenvalley_2019,zhang_bingohod_2022,li_nonlinearmodelling_2024,hitz-fasthisims_2025}.
In all cases, a suite of simulations with the correct non-Gaussian correlations
will be required to validate the analysis pipeline and evaluate covariance matrices.
We leave these tasks for future work; instead, in this initial study, we obtain an
expectation for the signal from a single simulated realization.

\subsection{Signal in Simulation}

We use a simulation to validate the cross-correlation method and provide an
expectation for the shape and amplitude of a signal. The \code{SkyLine} code
\citep{skyline_2023} is used to paint \HI intensity onto galaxy catalogs derived
by \code{UniverseMachine} \citep{unimachine_2023} from the $N$-body simulation
\code{MultiDark Planck 2} \citep{mdpl2_2012}. \code{SkyLine} models the halo \HI
mass function with the fitting formula and parameters derived from
\code{IllustrisTNG} by \citet{21cm_ingredients_2018}. That study of \HI in
\code{IllustrisTNG} also finds that $98\%$ of \HI mass is found in halos more
massive than $6.9\times10^9\ \mathrm{M_\odot} / h$. The \code{MultiDark} mass
resolution of $1.5\times10^9\ \mathrm{M_\odot} / h$ should thus be sufficient to
capture the bulk of the HI mass, but will not properly resolve the lowest-mass
halos containing HI. We leave it to future work to quantify the impact of these
missing halos on \tcm observables like the bispectrum of interest in this work.

A powerful feature of the \code{SkyLine} maps is that they are designed to share
the lightcone structure with the \code{AGORA} suite of simulated CMB observables
\citep{agora_2024}. The maps simulated in this way for the CHIME frequency band
thus possess the crucial features to capture the three-point correlation we aim
to measure: they are derived from an $N$-body simulation with sufficient mass resolution and therefore capture non-linear gravitational evolution as traced by HI; and they are correlated with a corresponding map of CMB lensing.

\begin{figure}
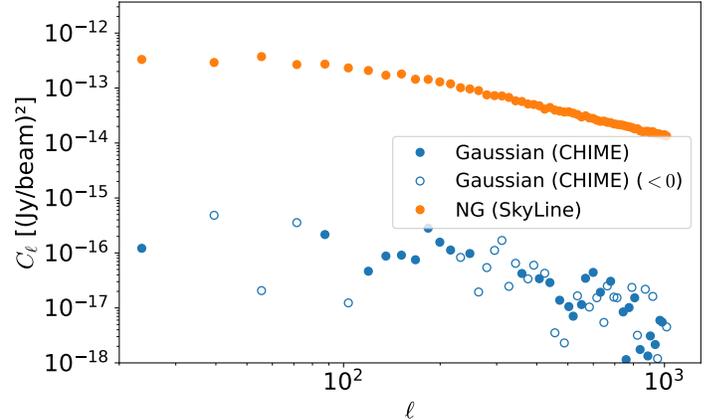

    \centfig[1.1]{raw_sim_angps_kappa_hi2_binned}
    \caption{
    Gaussian simulations of the \tcm signal (blue dots) do not capture the three-point correlation function, and thus predict a null cross-power spectrum of CMB
    lensing and \tcm variance along the line of sight 
    (Eq~\eqref{eq:xcorr-th}).
    Nonlinear simulations such as \texttt{SkyLine}, based on N-body simulations, are thus needed to predict our signal (orange dots).
    }
    \label{fig:sims-bispec}
\end{figure}

In Figure~\ref{fig:sims-bispec} we demonstrate this by computing the
angular cross-power spectrum of the simulated CMB lensing and \tcm line-of-sight
variance maps from \code{AGORA} and \code{SkyLine} respectively, and comparing it to the same
quantity estimated from \tcm and lensing maps derived from a gaussian
realisation and linear evolution. This higher-order signal is clearly present in
the $N$-body derived simulations but not in the gaussian ones.

%% file: data.tex
\section{Data}
\label{sec:data}

\subsection{CHIME \tcm Maps}

The CHIME maps used in this work are the same that were prepared for the
auto-spectrum analysis published in \citet{chime_ps_2025} and we refer the
reader to that paper for a detailed description of the observations and
processing that were used to produce them. A number of improvements to the
pipeline that was first presented in \citet{chime_collaboration_detection_2022}
have led to significantly cleaner maps.

Briefly, the maps are generated from a small subsample of $94$ nights from
2019 (out of $6$~years of recorded observations) that have been the focus of analysis efforts so
far. They are restricted to the $\sim600-700$~MHz band ($1.0 \lesssim z \lesssim
1.3$) that is relatively clean of RFI. Individual nights go through several
stages of calibration and flagging before a high-pass filter in delay (cutting
out $\tau < 200$~ns) is applied to suppress foreground power and its leakage to
higher delays due to uncalibrated instrumental response. The filtered nights are
averaged together to produce a co-added map of the sidereal day at every
frequency. In addition to the full co-add, nights are split into two partitions
(`even' and `odd') to be averaged separately into independent measurements of
the sky, which are useful for, e.g., internal cross-correlation or null tests.

A key step of the processing pipeline is to synthesise maps from the measured
visibilities, and this involves angular filtering. This first occurs via the selection of
baselines that are included in the synthesis. We exclude baselines that are
composed of feeds both within the same cylinder due to the sensitivity of these baselines to noise cross-talk and Milky Way foregrounds,
which sets the shortest baseline to $\sim22$~m -- effectively high-pass
filtering the maps. The map-making stage also filters out modes that are poorly
measured, based on a simple model for the observing geometry of the telescope
\citep{chime_collaboration_detection_2022}.

The CHIME maps are generated on a sky projection that corresponds to the
driftscan operation of the telescope and its cylindrical reflectors. They are
sampled evenly in right ascension (RA) along rings of constant declination that
are distributed equally in $\sin \theta_z$, where $\theta_z$ is the angle from
zenith. This projection is referred to as a ringmap. Although they measure
surface brightness, the solid angle of CHIME's interferometric beam is not
presently well calibrated, so the units we use in the maps are Jy/beam,
referenced to the flux of Cygnus A \citep{chime_collaboration_detection_2022}.

\begin{figure*}
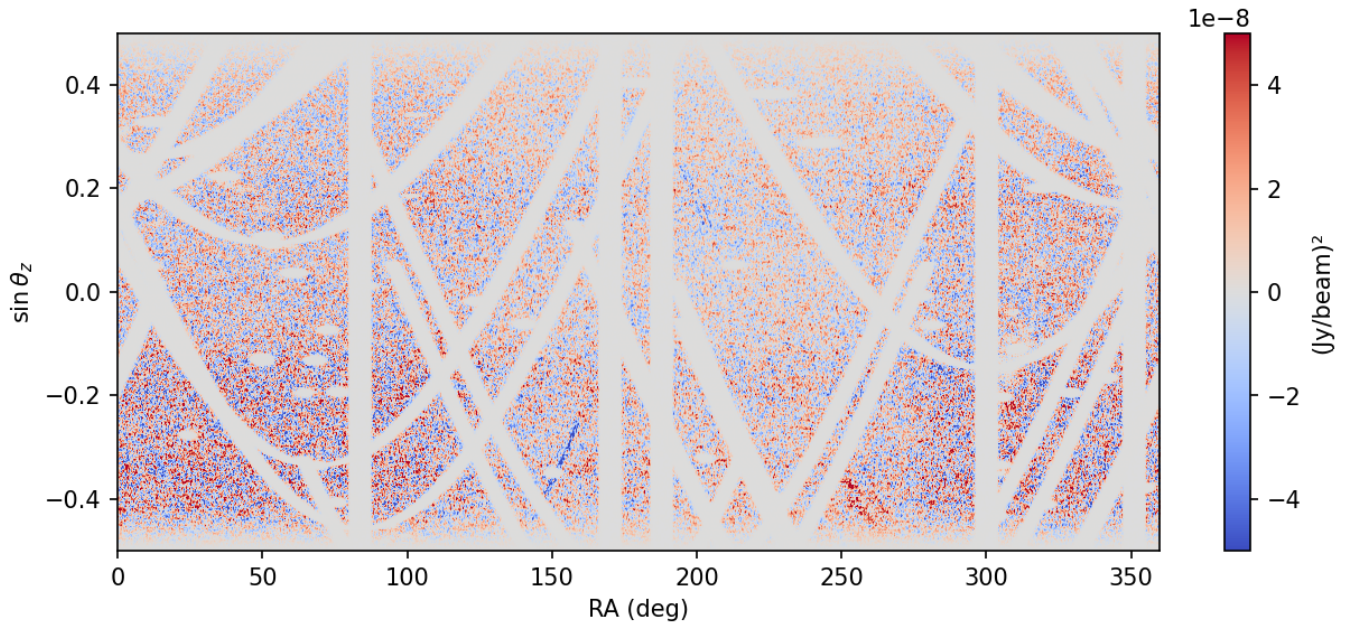

    \centfig[1.0]{masked_rmchime_psspec_srcmask}
    \caption{Map of line of sight variance estimated from the filtered CHIME
    data. The vertical axis is $\sin$ of the angle from zenith, a coordinate
    that corresponds to the Fourier conjugate of the telescope's North-South
    baseline grid. The locations of bright points sources are masked in a region
    around their meridian crossing, when they are directly overhead CHIME (small
    ellipses). For the brightest sources, we additionally mask the tracks they
    follow as they transit through the far side-lobes of the telescope beam
    (curved tracks), and all declinations at their transit time (vertical bars).
    The top and bottom of the map have been smoothly tapered. Residual
    contamination can be seen near the plane of the Galaxy and along tracks from
    sources outside the field that are not masked.
    }
    \label{fig:chime-var}
\end{figure*}

Regions of the map near bright point sources tend to have residual
contamination that survives the delay filter due the intensity of their emission
and the frequency dependence of the beam shape that imprints structure on their
spectra. Similar to what is done in \citet{chime_ps_2025}, the brightest sources
(above 60~Jy) are masked along tracks that follow their trajectory as they drift
through the far sidelobes of the instrument. Fainter sources (above 10~Jy) are
masked in a region around their transit corresponding to the main lobe of the
beam.

Finally, we restrict the extent of the map considered in this analysis to a
region $\pm30$\degree about zenith along the declination direction. Beyond this, the response of the beam begins
to roll off, but more problematically, aliasing becomes an issue
\citep{chime_collaboration_detection_2022}. At 650~MHz, the $\sim30$~cm shortest
separation between feeds is insufficient to Nyquist sample the sky from horizon
to horizon and causes aliasing of structures further than 32\degree from zenith,
motivating the choice of a $\pm30$\degree cutoff. To avoid introducing
ringing in subsequent transformations, a smooth taper is applied to the edges of
the mask. The final masked extent of the map is shown in
Figure~\ref{fig:chime-var}.

The $\pm30$\degree CHIME ringmap corresponds to approximately $13500\ \text{deg}^2$ of sky area. The point source mask described above removes about 24\% of this, leaving $\sim10300\ \text{deg}^2$ used in this analysis.

\subsubsection{Line of Sight Variance Estimation}

The quantity we aim to cross-correlate with lensing of the CMB is \tcm power
along the line of sight, computed from the CHIME multi-frequency maps. Starting
from CHIME ringmaps $d(\theta_j, \phi_k, \nu_n)$ in rings of constant
declination $\theta_j$, sampled at RA $\phi_k$, and frequencies $\nu_n$, we wish
to estimate the variance along the line-of-sight (the $\nu_n$ axis),
\begin{equation}
    v(\theta_j, \phi_k) = \langle \lv d(\theta_j, \phi_k, \nu_n) \rv^2 \rangle_\nu \ ,
\end{equation}

The measured map $d(\theta_j, \phi_k, \nu_n)$ includes noise, which will
introduce a bias if we just estimate the variance by squaring it. To mitigate
this issue, we estimate the variance by computing an internal cross-product of
maps from the two polarisations, labelled by $X$ and $Y$.\footnote{To be precise, the two maps correspond to CHIME's $XX$ and $YY$ visibilities, containing correlations of $X$ or $Y$ polarizations of each antenna. For brevity, we refer to these simply as $X$ and $Y$ in this work.} 
The \tcm signal is
unpolarised, so its power in both polarisations is correlated, but the noise in
these is uncorrelated, which will result in an unbiased estimator of the
variance (at least in the case of thermal noise). Each data point is accompanied
by an inverse variance weight $w(\theta_j, \phi_k, \nu_n)$ derived from the
fast-cadence thermal noise estimate measured during data acquisition
\citep{chime_collaboration_detection_2022}. These are used to weight the line of
sight average.

We estimate the cross-variance of the two maps as
\begin{equation}
    v = \langle d^X d^Y \rangle - \langle d^X \rangle \langle d^Y \rangle\ ,
\end{equation}
where the angle brackets denote a weighted mean using a combination
of the weights from both polarisations:
\begin{equation}
    w_n = \lp \sum_n \sqrt{w^X_n w^Y_n} \rp^{-1} \sqrt{w^X_n w^Y_n} \ .
\end{equation}

\subsection{Planck CMB Lensing}

The CMB lensing dataset used in the analysis is the baseline minimum variance (MV) map
from Planck PR3 \citep{collaboration_planck_2019}. We apply a Wiener filter to the
harmonic coefficients derived from the accompanying signal and noise power
spectra: $W_L = C_L/\lp C_L + N_L \rp$. The galactic region is excluded using
the provided mask. We do not explicitly apply any $\ell$-cut, but we restrict
the analysis to small patches of sky in a way that will be explained in the
following section.

%% file: method.tex
\section{Cross-correlation Procedure}
\label{sec:method}

To estimate the cross-correlation of the CHIME \tcm variance map with the Planck
lensing map, we must account for the particular geometry of the CHIME ringmap,
its partial sky coverage and the masked regions around bright point sources and
our galaxy. These factors prevent the reliable estimation of low-$\ell$ modes,
which are in any case suppressed by the spatial filtering of the CHIME maps, and
motivate a tiled flat-sky approach to the cross-correlation. Our chosen method
is summarised visually in Figure~\ref{fig:diagram}.

We divide the sky area selected from the CHIME map into square patches 10\degree
wide, a size that roughly balances mask avoidance and the inclusion of scales
close to the peak of the lensing spectrum. These are sampled on a $64\times64$
pixel grid. The number of patches, 128, is chosen to cover the entire field,
while minimising overlap between the patches. The pixel density roughly
matches $\ell_{max} = 2048$, chosen as sufficient to capture the smallest scales
in the CHIME map. The resulting tiling is featured in Figure~\ref{fig:diagram}.
Both maps are projected onto this tiling by first transforming them to spherical
harmonics and synthesising the small patches from these.

It is then straightforward to estimate the 2-d cross-power spectrum on these
flat patches via Fourier transform along each axis. Patches with $>1/3$ of their
area masked are excluded and a Blackman-Harris apodisation is applied along both
axes of the patch prior to the angular transform. Pairs of Fourier patches from
the CHIME and Planck maps are multiplied and averaged to compute the cross-power
spectrum. Signal to noise is expected to be low, so we choose to search for a signal
in the correlation function, where signal power will be concentrated in the
zero-lag bin. We take the inverse 2-d Fourier transform of the power spectrum and
bin it azimuthally in order to produce a 1-d cross-correlation function from these
two fields.

Our tiled cross-correlation approach
provides another benefit: randomly permuting the order of the patches from one
of the fields before computing the cross-correlation is an easy way to produce
many null realisations from the data. Permuting the patches in this way removes
the signal correlation but preserves sources of contamination such as noise and
foreground residuals in the resulting correlation functions (the exception would
be sources of contamination that are actually correlated between the \tcm
variance and lensing fields). We use this trick extensively to estimate the
covariance of the total effect of these noise contributions from the data itself.

%% file: results.tex
\section{Results}
\label{sec:results}

\begin{figure}[tb]
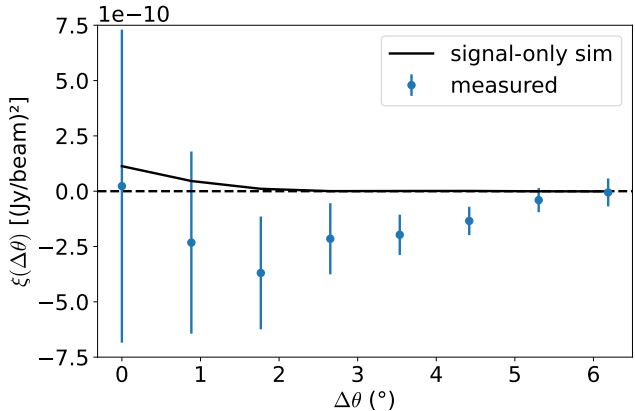

    \centfig[1.0]{cf1d_1sig_fitchime_psspec_srcmask}
    \caption{The measured correlation function for Planck lensing and CHIME
    variance maps (blue points with error bars) is consistent with noise. 
    Indeed, the chi-squared with respect to the null hypothesis (i.e. no
    signal model) is $\chi_\nu^2=0.90$, with a PTE of $0.51$.
    This includes the non-zero correlation between data points.
    }
    \label{fig:data-corrfunc}
\end{figure}

Figure~\ref{fig:data-corrfunc} shows the result of applying the procedure
described in the previous section to the CHIME variance and Planck CMB lensing
maps. Also shown are the expected signal from simulations (without noise), as
well as the level of residuals measured from null permutations of the data. It is
clear that this is a non-detection, with the expected signal amplitude a factor
of a few below the noise. We can quantify the signal to noise (S/N) by
calculating the uncertainty on a fit to the data given the measured covariance.
We model the signal with a single amplitude parameter scaling the shape of the
expected signal from the simulations:
\begin{equation}
    m(\theta_i) = a t(\theta_i)\ ;\ \vec{m} = T \vec{a}\ ,
\end{equation}
where $t(\theta_i)$ is the signal template obtained from the simulations and
evaluated at angular separations $\theta_i$. This is cast on the right as a
design matrix $T$ and parameter vector $\vec{a}$ (containing a single element).

\begin{figure}[tb]
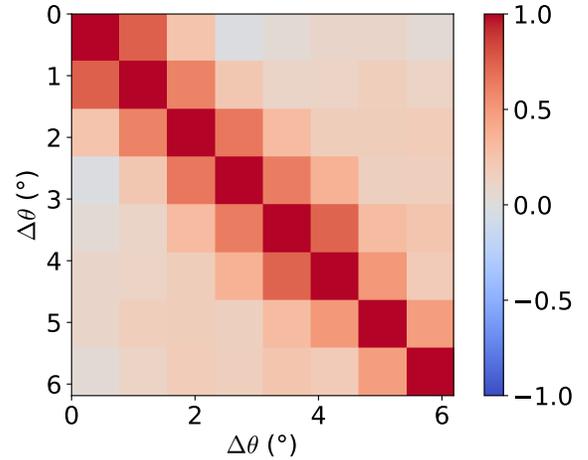

    \centfig[0.9]{cf1d_cov}
    \caption{
    The correlation matrix between the samples in
    Figure~\ref{fig:data-corrfunc} shows that they are not independent.
    }
    \label{fig:data-cov}
\end{figure}

The covariance of the measurement is computed from $N=128$ random permutations.
For each $k^\mathrm{th}$ permutation, we evaluate a correlation function
$p_k(\theta_i)$ and compute the covariance by averaging over them:
\begin{equation}
    \label{eq:null-cov}
    C(\theta_i, \theta_j) = \frac{1}{N}\sum_k^N p_k(\theta_i) p_k(\theta_j)\ .
\end{equation}
The correlation between individual samples is significant, as can be seen in
Figure~\ref{fig:data-cov}. Using this quantity, we evaluate
\begin{equation}
    \chi^2 = \lp\vec{d} - T\vec{a}\rp^T C^{-1} \lp\vec{d} - T\vec{a}\rp\ ,
\end{equation}
given the data vector $\vec{d}$. The solution that minimises $\chi^2$ is
\begin{equation}
    \vec{a} = \lp T^T C^{-1} T \rp^{-1} T^T C^{-1} \vec{d} \ ,
\end{equation}
with inverse variance
\begin{equation}
    \label{eq:sn}
    \sigma_a^{-2} = T^T C^{-1} T \ .
\end{equation}
This is the uncertainty on the measurement in units of the expected signal
amplitude, which we interpret as the S/N achievable given the covariance of the
data, under the assumption that the true signal is well represented by the
simulations -- S/N~$\sim \sigma_a^{-1}$. For the datasets considered here
(Figure~\ref{fig:data-corrfunc}), we find S/N~$\sim 0.2$. The fit to the
measurement results in a reduced $\chi^2_\nu=1.03$ and a corresponding PTE of
0.41. If we assume no signal model (i.e. a model where the data is just noise),
we find $\chi_\nu^2 = 0.90$ and a PTE of 0.51, and we conclude that given the
current errors we cannot distinguish between a signal and the null hypothesis.
We explore how adding more data or relaxing some of the imposed constraints may
improve detectability in the future in Section~\ref{sec:forecasts}. In the
remainder of this section we compare the measurement to a null test and
simulations in order to investigate what is contributing to the observed
residuals.

\subsection{Comparison to Null Test and Simulation}

We can use the even and odd partitions of the CHIME observations to compute a
variance map from their difference rather than their sum and run that through
the cross-correlation procedure. The effect of the even-odd difference is to
null components of the map that are unchanging from one sidereal day to the
next, i.e. the \tcm and foreground sky. Residuals above the expected thermal
noise level point to sources of contamination that change in time, such as RFI
events or errors in the instrumental calibration.

\begin{figure*}[htb]
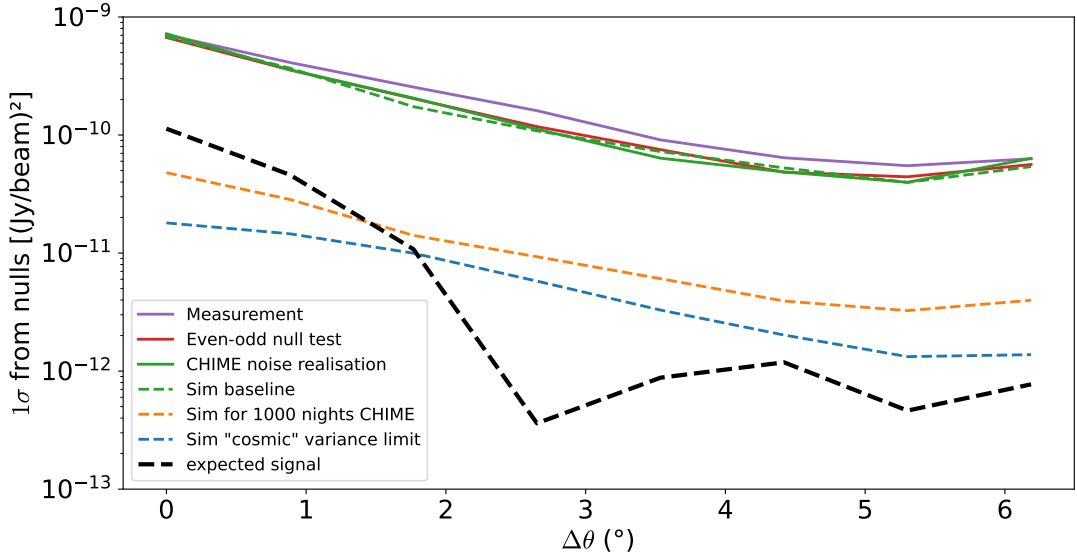

    \centfig[0.8]{data_corrfunc_var}
    \caption{
    We estimate the measurement noise in multiple ways, finding good consistency.
    From the data alone, we correlate \tcm tiles with mismatched CMB lensing tiles, obtaining the standard deviation of permutations (purple).
    We also replace CHIME data with the difference between odd and even nights (red) or with a CHIME noise realization (solid green).
    Finally, we replace the CHIME data by a noise realization plus the negligible signal from SkyLine and the Planck data by a simulated noise and signal from Agora (dashed green).
    Given this consistency, we can confidently scale the thermal noise floor  
    with an increased number of CHIME
    observations (dashed orange), and compare to the expected signal amplitude (dashed black).
    The result is a factor of about two larger than the sample variance limit (noiseless, dashed blue)
    due to the finite number of modes measured after all of the
    required data cuts. 
    Note that in all cases, the restricted frequency band
    and strict delay filter used in this analysis are maintained.
    }
    \label{fig:corrfunc-var}
\end{figure*}

Figure~\ref{fig:corrfunc-var} shows the standard deviation for the correlation
function measured across null permutations (Equation~\ref{eq:null-cov}) for the
co-add of CHIME maps, even-odd difference and a realisation of gaussian noise at
the expected thermal level. The noise realisation has been propagated through
the analysis pipeline including the masking and filtering steps. There we see
that the even-odd difference is consistent with the expectation for thermal
noise, and that the measurement is slightly elevated compared to those. This
indicates that there exists some residual contamination above the noise in the
CHIME data. We do see hints of residual structures in the map of
Figure~\ref{fig:chime-var}, which may explain this excess. However, overall the
cleaned maps appear to be dominated by thermal noise and we do not attempt to
optimise the mask any further given the current sensitivity.

Finally, the green dashed curve in Figure~\ref{fig:corrfunc-var} is the
covariance estimated on simulated maps of CMB lensing and \tcm, including
gaussian noise with amplitude corresponding to the Planck and CHIME maps. It
closely follows the case where a noise realisation was substituted for the CHIME
data, which validates that the simulation pipeline is generating products that
are consistent with the data.

%% file: forecasts.tex
\section{Forecasts for Larger Data Volumes}
\label{sec:forecasts}

The simulations that were generated to compare to the measurement we performed
need not be limited by the same constraints as the data currently is. In this
section we use them to produce forecasts for future detectability of the signal.
We will limit ourselves to specific extensions of the datasets that are
straightforward to evaluate from the products available to us and test the
effect of improved noise within either the \tcm or CMB lensing observations.

The data presented here is just a small fraction of what CHIME has collected to
date -- in terms of integration time, but also bandwidth, sky coverage and modes
lost to very aggressive foreground cleaning. Efforts to overcome the
challenges that impose these constraints continue.
A direction in which progress is expected in the near term is the processing
of additional CHIME observations to add to the maps, which will lower the 
thermal noise floor. This motivates a forecast scenario in which data cuts 
for cleaning are assumed to remain the same, but the
total number of observations is increased by a factor of 10, to 1000 nights.

The background fluctuations for this case are displayed in
Figure~\ref{fig:corrfunc-var}. Also shown is the background for simulations that
contain signal only, which are due to sample variance from the limited number of
modes that survive the data cuts and represents the best possible measurement
for this configuration. The 1000 nights of CHIME data are within a factor of
$<2$ of this best case across the board. S/N, as defined in
Equation~\ref{eq:sn}, is $\sim 3$ compared to $\sim11$ for the noiseless case.
S/N figures for all cases considered in this work are summarised in
Table~\ref{tab:sn-list}.

\begin{figure}[htb]
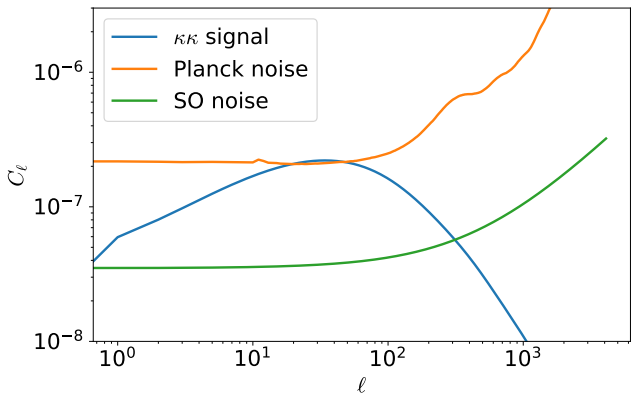

    \centfig[1.0]{so_noise_spec}
    \caption{To assess the impact of the lensing noise on the measurement, we
    also substitute the noise power spectra forecasted for the SO LAT in
    \citet{lat_forecast_2019} (green) in place of the Planck noise (orange). The fiducial lensing
    power spectrum is also shown for comparison (blue).
    Note that SO will not provide
    the same sky overlap with CHIME as Planck does.
    }
    \label{fig:so-noise}
\end{figure}

Next, in addition to the extended CHIME integration time we consider a scenario
of improved noise in the CMB lensing reconstruction, at the level forecast by
the Simons Observatory (SO) \citep{lat_forecast_2019}, see
Figure~\ref{fig:so-noise}. As opposed to the previous scenario, this does not
represent a configuration that will realistically be available in the near
future, since the overlap between SO lensing maps in the southern hemisphere and
the CHIME maps will be much less than what is possible with Planck.
Nevertheless, we find a S/N~$\sim4$ in this case, a significant improvement that
reflects the fact that in a cross-correlation noise terms are multiplied and
thus improvements from either side contribute regardless of their intrinsic S/N.

Although the S/N forecast assuming the current foreground mitigation measures is
modest, the fact that a first detection appears possible should be encouraging,
and motivate continued efforts to relax the necessary cuts and unlock access to
the bulk of the signal available to CHIME. It is worth emphasizing that the
effective volume of data used in this work is a small fraction of what has been
measured. We anticipate that continued improvements in removing contaminants
from the data will enable a relaxing of constraints beyond what was considered
in this section. In particular, reducing the cut-off of the delay filter could
substantially improve detectability.

\begin{table}
\begin{tabularx}{1.0\columnwidth}{l r}
    \textbf{Configuration} & \textbf{S/N} \\
    \hline
    Measured noise: CHIME x Planck & 0.2 \\
    Simulated noise: CHIME x Planck & 0.2 \\
    Simulated: CHIME 1000 nights x Planck & 3 \\
    Simulated: CHIME 1000 nights x SO-like & 4
\end{tabularx}
\caption{S/N for cross-correlation measurement and simulations that vary
the noise properties of the CHIME and lensing maps (see Equation~\ref{eq:sn} for
definition of S/N).
}
\label{tab:sn-list}
\end{table}

%% file: conclusion.tex
\section{Conclusion}

Cross-correlation with external surveys has proven to be a powerful tool in the
development of \HI intensity mapping as a cosmological probe, notably using
measurements of galaxy positions and the Lyman-$\alpha$ forest from eBOSS. At
higher redshifts however, targets complementary to the \tcm observations become
more sparse. Weak lensing of the CMB could fill this gap as it is a direct
measure of the matter density projected along the line of sight, with a broad
kernel peaking around redshifts $z\sim1-10$. 
Measuring this cross-correlation is
made difficult by the extremely bright \tcm foregrounds that occupy the same
slowly-varying line of sight modes as the lensing signal.

In this work we have implemented a position-dependent power spectrum analysis to
attempt to measure the squeezed bispectrum in the cross-correlation of line of
sight variance of \tcm maps from CHIME and the lensing convergence map derived
by Planck. This method makes up for the modes lost to foregrounds by targeting
the coupling due to non-linear gravitational evolution of small-scale
fluctuations in the \tcm field with the large-scale modes measured by lensing.
Using simulations that capture this non-linear effect, we forecast
the expected signal level and compare to what we measure in the data. We find
that given the current noise level in the CHIME data, the expected signal to
noise is only $\sim 0.2$. Indeed, the measured correlation function is
statistically consistent with noise, with a $\chi^2$ with respect to the null
hypothesis of $0.90$ (PTE~$\sim0.51$). We forecast that increasing the number of
observations in the CHIME maps by a factor of 10 would reduce thermal noise
to the level of a $3\sigma$ detection. This does not account for other sources
of contamination such as foreground and RFI residuals that may become
significant with suppressed noise.

The CHIME data used in this analysis are restricted in the number of nights of
observations included and the aggressive foreground cut that is applied. We
anticipate both of these constraints can be relaxed in the near future. The
observations used in this work represent less than $1/10$th of the
integration time available to be analysed, and inclusion of the full dataset
is a priority for the next stages of the CHIME effort. Improvements to
calibration and flagging continue to be developed and tested, which will result
in less contamination leaking to high delays, and enable a less
harsh delay filter used to suppress foreground power. Given the signal level
measured in simulations, one can be optimistic that continued refinement of
the CHIME processing and analysis can lead to a detection of the bispectrum signal.

In this work we have not explored the possibility of performing this
cross-correlation with other \tcm datasets, but there exist exciting candidates.
In particular, observations from the southern hemisphere would benefit from
overlap with measurements of CMB lensing that have lower noise levels than those
of Planck. ACT has produced maps that are signal-dominated over a range of
scales \citep{madhavacheril_atacama_2023,qu_act_ps_2024} and SO is forecasted to
improve reconstructed CMB lensing signal-to-noise in the near term (Figure~\ref{fig:so-noise}). 
In the south, HERA has been observing the \tcm sky over a large area at redshifts corresponding to the epoch
of reionisation \citep{hera_ps_2022}, and SKA and its precursors will provide
next-generation maps from Australia and South Africa. Adapting the simulations
that were used here to these observational configurations is a logical next step
that would provide insight into the possibilities for this measurement in the
future, in conjunction with further study of the theoretical modeling necessary
to interpret it for cosmology.

%% file: ack.tex
\section{Acknowledgements}


This work received support from the Department of Energy, Laboratory Directed Research and Development program at SLAC National Accelerator Laboratory, under contract DE-AC02-76SF00515.

We thank the Dominion Radio Astrophysical Observatory, operated by the National
Research Council Canada, for gracious hospitality and expertise. The DRAO is
situated on the traditional, ancestral, and unceded territory of the syilx
Okanagan people. We are fortunate to live and work on these lands.

CHIME is funded by grants from the Canada Foundation for Innovation (CFI) 2012
Leading Edge Fund (Project 31170), the CFI 2015 Innovation Fund (Project 33213),
and by contributions from the provinces of British Columbia, Québec, and
Ontario. Long-term data storage and computational support for analysis is
provided by WestGrid\footnote{\url{https://www.westgrid.ca/}},
SciNet\footnote{\url{https://www.scinethpc.ca/}} and Digital Research Alliance
of Canada\footnote{\url{https://www.alliancecan.ca/}}, and we thank their
staff for flexibility and technical expertise that has been essential to this
work, particularly Martin Siegert, Lixin Liu, and Lance Couture.

Additional support was provided by the University of British Columbia, McGill
University, and the University of Toronto. CHIME also benefits from the Natural
Sciences and Engineering Research Council of Canada (NSERC) Discovery Grants to
several researchers, funding from the Canadian Institute for Advanced Research
(CIFAR), and the FRQNT Centre de Recherche en Astrophysique du Québec (CRAQ).
Specific funding references from NSERC that supported this work are 569654 and
RGPIN-2023-05373. This material is partly based on work supported by the NSF
through grants (2008031), (2510770) and (2510771).

Nickolas Kokron acknowledges support from the Bershadsky Fund and the Fund for
Natural Sciences of the Institute for Advanced Study.

This work was based on observations obtained with
Planck\footnote{\url{http://www.esa.int/Planck}}, an ESA science mission with
instruments and contributions directly funded by ESA Member States, NASA, and
Canada.

The authors gratefully acknowledge the Gauss Centre for Supercomputing
e.V.\footnote{\url{www.gauss-centre.eu}} and the Partnership for Advanced
Supercomputing in Europe (PRACE\footnote{\url{www.prace-ri.eu}}) for funding the
MultiDark simulation project by providing computing time on the GCS
Supercomputer SuperMUC at Leibniz Supercomputing Centre (LRZ, www.lrz.de). The
Bolshoi simulations have been performed within the Bolshoi project of the
University of California High-Performance AstroComputing Center (UC-HiPACC) and
were run at the NASA Ames Research Center.

Some of the computing for this project was performed on the Sherlock cluster at Stanford University. We
would like to thank Stanford University and the Stanford Research Computing
Center for providing computational resources and support that contributed to
these research results.